# Heat conduction of single-walled carbon nanotube isotope-superlattice structures: A molecular dynamics study

Junichiro Shiomi and Shigeo Maruyama[*]

Department of Mechanical Engineering, The University of Tokyo
7-3-1 Hongo, Bunkyo-ku, Tokyo 113-8656, Japan
*Corresponding author: Tel & Fax: +81-3-5800-6983.
E-mail address: maruyama@photon.t.u-tokyo.ac.jp

Heat conduction of single-walled carbon nanotubes (SWNTs) isotope-superlattice is investigated by means of classical molecular dynamics simulations. Superlattice structures were formed by alternately connecting SWNTs with different masses. On varying the superlattice period, the critical value with minimum effective thermal conductivity was identified, where dominant physics switches from zone-folding effect to thermal boundary resistance of lattice interface. The crossover mechanism is explained with the energy density spectra where zone-folding effects can be clearly observed. The results suggest that the critical superlattice period thickness depends on the mean free path distribution of diffusive-ballistic phonons. The reduction of the thermal conductivity with superlattice structures beats that of the one-dimensional alloy structure, though the minimum thermal conductivity is still slightly higher than the value obtained by two-dimensional random mixing of isotopes.

## I. Introduction

The ever-expanding expectations for single-walled carbon nanotubes (SWNTs) [1] include applications for various electrical and thermal devices due to their remarkable electrical and thermal properties [2, 3]. On considering the actual applications, one of the essential tasks is to characterize the thermal properties certainly for thermal devices and also for electrical devices since they limit the affordable amount of electrical current through the system. In actual applications, SWNTs may have impurities, defects and junctions, which alter the heat conduction. In general, these nanoscale impurities, having scales comparable to the phonon mean free path $l$ of the system, are expected to strongly influence the thermal properties of bulk materials. For SWNTs, due to the quasi-one-dimensional structure and the expected long $l$, the impact and its scale dependence of the impurity-effects may differ considerably from those of the three-dimensional materials. In our previous molecular dynamics study, isotope-effects on the thermal conduction were investigated by randomly mixing $^{13}$C isotopes in a finite length $^{12}$C-SWNT with chirality of (5, 5) [4]. The results showed that the effective thermal conductivity decreases monotonically as the number ratio of $^{13}$C atoms increases. Interestingly, the isotope-effect was found to be considerably weaker than that in diamonds. This led us to speculate that the phonon impurity scattering may be restricted in SWNTs due to the one-dimensional confinement of phonons.

In the current work, we proceed with the characterization of heat conduction of SWNTs subjected to nanoscale intrinsic thermal resistances. Here, in order to reduce the complication due to the disorderliness of impurities in the previous work [4], we investigate the heat conduction of SWNTs altered by systematically placed thermal resistances with well-defined scales. To this end, the superlattice structures were constructed by periodically connecting SWNTs with different masses. There have been a considerable number of studies devoted to characterize thermal properties of superlattice structures in three-dimensional materials. Being motivated by thermoelectric device applications, the main purpose of these studies has been to utilize the superlattice structures to reduce the effective thermal conductivity $\lambda$ in the direction normal to the interfaces. The challenge has been to beat $\lambda$ of alloys (alloy limit) which classically, among binary crystal materials, has been considered to possess the minimum thermal conductivity. Experimental and theoretical studies of the cross-plane thermal conductivity and its dependence on the period thickness have been reported for thin film superlattices [5-12] and recently for nanowire superlattices [13-15]. The most common experimental observation has been the reduction of thermal conductivity on shortening the period thickness $\Delta z$ [5] and hence extensive theoretical studies have been directed to reproduce such trend [6]. On the other hand, experiments of short-period superlattices revealed the existence of the minimum thermal conductivity for a period thickness of about 5 nm [7, 8]. This was explained by a study using molecular dynamics models, where the minimum thermal conductivity was observed for a period thickness of several monolayers [9].

The appearance of the minimum thermal conductivity can be understood in terms of two competing mechanisms that yield opposite dependence of $\lambda$ on $\Delta z$ [10]. One is the zone-folding effect due to new periodicity imposed by the superlattice structure. The reduction of the group velocity and enhancement of Umklapp scattering at the imposed zone-boundaries (mini-Brillouin zone boundaries), which appear as the gaps in the dispersion relations (mini-bandgaps), attenuate the thermal conductivity [11, 12]. The effect becomes stronger as $\Delta z$ increases since the number of folds is proportional to the period thickness. The other mechanism is the thermal boundary resistance (TBR) at the superlattice interfaces due to phonon reflection and scattering. The TBR effect increases with the number of interfaces per length and hence decreases with $\Delta z$. The reduction of $\lambda$ attributed to TBR typically saturates on reducing $\Delta z$ bellow the characteristic length of the ballistic phonon transport due to the phonon tunneling effects i.e. the transboundary ballistic transport of phonons through TBRs. Simkin and Mahan [10], in their quantum mechanical model, implemented the crossover of the two competing physics by using the imaginary wavevector. The results led them to suggest that the critical period thickness $\Delta z_{cr}$ roughly scales with the phonon mean free path $l$.

Following the above idea of crossover, if the critical period thickness $\Delta z_{cr}$ is characterized with the mean free path $l$, $\Delta z_{cr}$ of an SWNT should be considerably large due to the expected long $l$. In our previous MD study on the length effect of the thermal conductivity of a pure (5, 5) SWNT at room temperature, the ballistic transport of phonons was observed up to at least a few micrometers [16].

Furthermore, the theoretical calculation of Mingo and Broido [17] suggests that the ballistic limit may exceed tens of micrometers. Therefore, when superlattice SWNTs are less than hundreds of nanometers long, which is a realistic length scale in many applications, the entire system is expected to be within the phonon tunneling range and the zone-folding effect would dominate over the TBR effect. On the contrary, we demonstrate that the thermal conductivity of superlattices can take a minimum value even in much shorter system well in the ballistic phonon transport regime.

As simulations of superlattices in three-dimensional materials inevitably suffer from uncertainties caused by simplifications of the boundary scattering and anisotropic effects, SWNTs would be the ultimate case where the theories can be applied without major simplifications due to the quasi-one-dimension confinement of phonons. The synthesis of SWNTs from $^{13}$C isotopes source has been demonstrated to be possible [18]. As the recent outstanding development of SWNT growth techniques using chemical vapor deposition in well controlled conditions, it may be possible to experimentally pattern superlattice structures in the near future.

**II. Molecular Dynamics Simulations**

Molecular dynamics (MD) simulations utilize Brenner potential [19] with a simplified form [20] to express interaction between carbons. This potential function has been reported to be able to describe variety of small hydrocarbons, graphite and diamond lattices. The formulation of the potential function is based on the covalent-bonding treatment developed by Tersoff [21]. The total potential energy of the system is expressed as

$$E = \sum_i \sum_{j(i<j)} \left[ V_R(r_{ij}) - B_{ij}^* V_A(r_{ij}) \right], \quad (1)$$

where $V_R(r)$ and $V_A(r)$ are repulsive and attractive force terms, which take the Morse type form with a certain cut-off function. $B^*_{ij}$ represents the effect of the bonding condition of the atoms. As for the potential parameters, we employ the set that was shown to reproduce the force constant better (table 2 in [19]). The velocity Verlet method was adopted to integrate the equation of motion with the time step of 0.5 fs. Simulations were done for armchair SWNTs with chiral index of (5, 5), which gives the radius of about 0.7 nm.

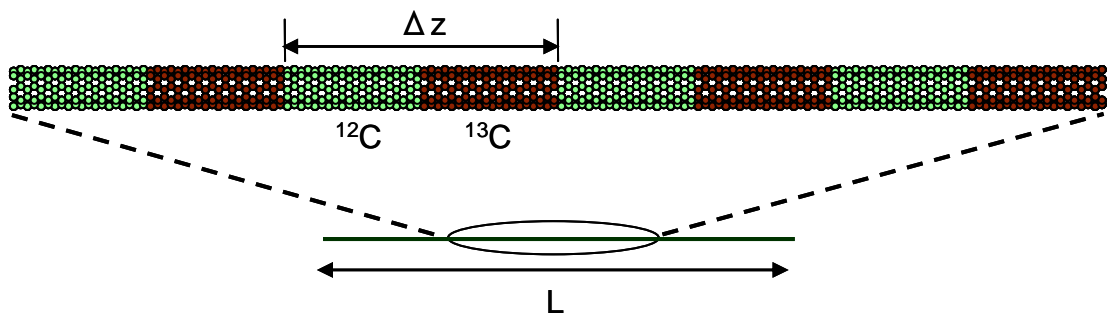

Fig. 1. (Color online) $^{12}$C/$^{13}$C (5, 5)-SWNT superlattice with period thickness $\Delta z$. The $^{12}$C/$^{24}$C system has the identical configuration.

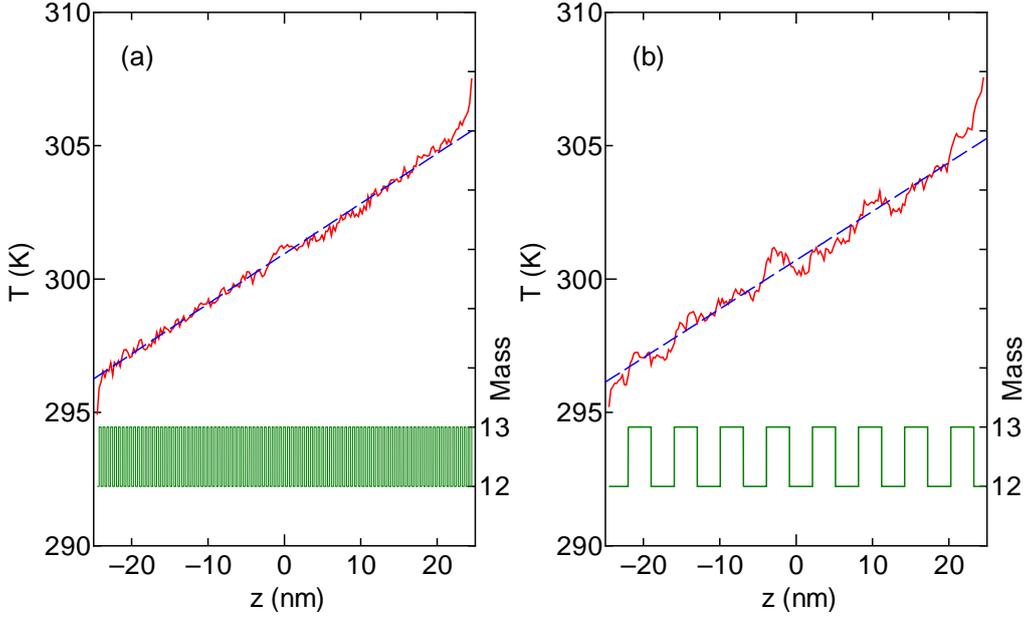

Fig. 2. (Color online) Temperature profiles on controlling the phantom atoms at different temperature 290 K ($z=-25$ nm) and 310 K ($z=25$ nm) with the relative mass profiles of $^{12}$C/$^{13}$C SWNT at the bottom. (a) $\Delta z=0.5$ nm, (b) $\Delta z=5.0$ nm.

Superlattice structures were constructed by alternately connecting (5, 5) SWNTs with different masses. Fig. 1 denotes an isotope superlattice with cells of $^{12}$C and $^{13}$C with a certain period thickness $\Delta z$. Simulations for $^{12}$C/$^{24}$C system were also carried out for clearer demonstration of the influence of $\Delta z$ on the heat conduction. Although, the former case is more realistic, as shown later, the fluctuations severely influence the statistics even with a considerable number of sampled data. Therefore, the latter case with higher mass ratio serves to highlight the targeted phenomenon by enhancing the signal to noise ratio.

The methodology of the thermal conductivity measurement follows our previous reports [22, 23]. The temperature gradient was applied by using the phantom atoms placed on both ends of an SWNT. Once a quasi-stationary temperature profile is achieved, $\lambda$ can be calculated from the temperature gradient and the heat flux $q$ obtained from energy budgets of phantom atoms using the Fourier's law, $q = Q/A = -\lambda \partial T/\partial z$. Here, the cross sectional area $A$ is defined as the area of a hexagon dividing a close packed bundle of SWNTs: $A = 2\sqrt{3}(d/2 + b/2)^2$, where $b$ is van der Waals thickness 0.337 nm. Although, the definition of $A$ is arguable, the intension here is simply not to exaggerate the value of $\lambda$ by choosing the definition which would result in relatively small $A$.

Simulations were performed with different $\Delta z$ and SWNT length $L$. Before imposing the temperature gradient, the average temperature of the system was set at 300 K with the auxiliary velocity scaling control. Measurements were started when the quasi-stationary state was achieved by the phantom thermostats maintained at 290 K ($z=-L/2$) and 310K ($z=L/2$). The measured data were averaged over 10-20 ns. Within the explored parameter space, the time-averaged temperature profile remained quasi-linear. As seen in the temperature profiles for $\Delta z=0.5$ nm [Fig. 2 (a)] and 5.0 nm [Fig. 2 (b)], the temperature profiles exhibit negligibly small bumps with the lengthscale corresponding to $\Delta z/2$. Heat fluxes added to and subtracted from the phantom thermostats at the tube-ends remained almost constant as those in the simulations of pure $^{12}$C-SWNTs [22, 23].

### III. Results and Discussions
#### A. Minima in thermal conductivity of $^{12}$C/$^{13}$C superlattices

The result of the simulations for $^{12}$C/$^{13}$C SWNT superlattice at room temperature shows that $\lambda$ takes a minimum value at $\Delta z_{cr} \sim 5$ nm (Fig. 3). The value is apparently much smaller than the expected $l$ of SWNTs. Although, the mean free path of the SWNT superlattice may well be shorter than that of a pure $^{12}$C-SWNT, the reduction of the mean free path should be limited to at most 50% as estimated by the reduction of the thermal conductivity.

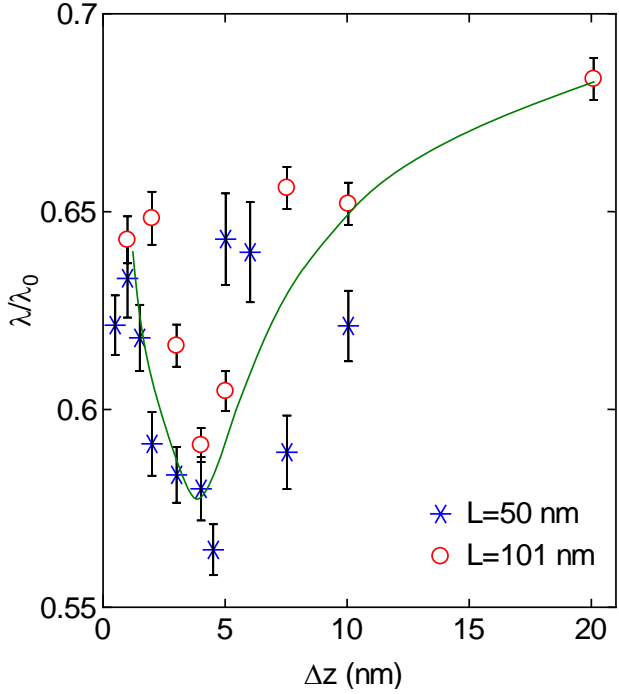

Fig. 3. (Color online) Dimensionless effective thermal conductivity for $^{12}C/^{13}C$ system and a range of period thickness $\Delta z$. Thermal conductivity for the pure $^{12}C$-SWNT $\lambda_0$ is 310 W/mK and 370 W/mK for $L$ = 50 nm and 101 nm, respectively. The line is drawn only for eye. The error bars are calculated based on the fitting residuals of the temperature profiles.

Let us now consider in terms of the contribution of individual phonons to heat conduction i.e. $\sum_i C_i v_i l_i$, in the regime with dominant influence of TBR ($\Delta z > \Delta z_{cr}$). We assign phonons into three categories with different contribution to $\lambda$; (1) fully ballistic phonons ($l_i \geq L$) which have $\Delta z$-invariant contribution to the conduction due to the phonon tunneling effect, (2) diffusive-ballistic phonons with $0 < l_i < L$, whose contribution to $\lambda$ is $\Delta z$-dependent and (3) stationary phonons with negligible contribution to heat conduction due to the small product of group velocity and mean free path, $v_i l_i$. The phonons in the second category are the only ones with noticeable contribution to the dependence of $\lambda$ on $\Delta z$. Therefore, if there is any mean free paths that should scale with $\Delta z_{cr}$ that would be the local mean free path of the diffusive-ballistic phonons. Yet, in MD simulations and also in reality, unlike the simple theoretical models with a single scale of mean free path, $l$ has a certain distribution and there is no assurance that it takes a simple form. Hence, the precise scale of $\Delta z_{cr}$ should depend on the distribution function of $l_i$ weighted by $C_i v_i$.

Thermal conductivity of an SWNT is known to exhibit distinct length effect [16, 22, 23], where $\lambda$ increases with the nanotube length. On varying $L$ from 50 nm to 101 nm, $\lambda$ of the pure $^{12}C$-SWNT ($\lambda_0$) increases from 313 W/mK to 370 W/mK due the length-effect [16, 22, 23]. Accordingly, $\lambda$ of $^{12}C/^{13}C$ SWNT superlattices increase with $L$ for the entire range of $\Delta z$. On the other hand, as seen in Fig. 3, $\Delta z_{cr}$ appears to be insensitive to the change of $L$. This is consistent with the above discussion that the fully ballistic phonons added by lengthening the nanotube simply increases the offset ($\Delta z$-independent) thermal conductivity and have minor influence to the dependence of $\lambda$ on $\Delta z$.

In our previous MD studies on the length effect of the SWNT thermal conductivity, the heat conduction exhibited a strong one-dimensional nature in the current length regime [16, 22, 23]. Assuming the SWNT superlattice to be a one-dimensional system, the superlattice with the smallest $\Delta z$ which consists of two monolayer unit cells can be considered as an 'alloy'. In this sense, the current results can be understood as a case where the attenuation of the thermal conductivity with superlattice structures beats the alloy limit. However, it should be noted that the minimum thermal conductivity is still slightly higher than $\lambda/\lambda_0$=0.53 ($\lambda/\lambda_0$=0.07 in case of $^{12}C/^{24}C$ system) obtained by randomly mixing $^{13}C$ atoms into a pure $^{12}C$-SWNT with a number ratio of 50 %.

### B. Model superlattice system with $^{12}C/^{24}C$

As seen in Fig. 3, even by sampling for 10-20 ns (20-40 million time steps), the statistics seriously suffer from fluctuations. In order to highlight the influence of superlattice structures and validate the above analyses, simulations were carried out for the $^{12}C/^{24}C$ system ($L$=50 nm). Consequently, as denoted with asterisks in Fig. 4, data are much smoother than in the previous case. Again, the minimum $\lambda$ is observed but this time the picture is more distinct. Larger mass difference results in smaller average $\lambda$ [$\lambda(\Delta z = \Delta z_{cr})$=30 W/mK] compared with the previous case. Correspondingly, $\Delta z_{cr}$ is reduced to 1.5 nm, which can be understood in terms of the narrowing of the phonon tunneling regime i.e. the reduction in mean free path of diffusive-ballistic phonons with $0 < l_i < L$ due to the enhancement of the impact of each lattice interface.

Fig. 4 also shows the temperature dependence of $\lambda$. On decreasing the temperature from 300 K to 50 K, mean free paths of the diffusive-ballistic phonons are expected to become longer. An apparent consequence is the increase of $\lambda$ due to the attenuation of the thermal phonon scattering (normal and Umklapp scatterings). In the current case, thermal conductivity of a pure $^{12}C$-SWNT increases from $\lambda_0$=310 W/mK (300 K) to $\lambda_0$=1167 W/mK (50 K).

Correspondingly, $\Delta z_{cr}$ is shifted upward on the horizontal axis and as a consequence $\lambda$ decreases monotonically without crossing the crossover limit in the entire range of $\Delta z$. The trend indicates that the TBR effect is small, which can be attributed to dominant ballistic phonon transport at low temperature for the current system size.

## C. Observation of phonon dispersion relations for superlattices

The simulations of the $^{12}C/^{24}C$ system enable us to clearly observe the zone-folding effect in the dispersion relations. The dispersion relations can be obtained from MD simulations by calculating the 2D-Fourier spectra of the time history of the 1D velocity field along an SWNT. Here, spectra are presented in terms of the energy density in $(\omega, k)$-space

$$E(\omega,k) = \frac{1}{3n}\sum_{\alpha}^{n}\sum^{3}\left|\frac{1}{N}\int v_{\alpha}(z,t)\exp(ikz-i\omega t)dtdz\right|^{2},$$
$$(\alpha = r, \phi, z) \qquad (2)$$

where $N$ and $n$ are the number of atoms in the longitudinal ($z$) direction (the number of unit cells in the nanotube) and number of atoms in a unit cell, respectively. The velocity vector is projected to the local cylindrical coordinates ($r, \phi, z$) denoted by the subscript $\alpha$ in Eq. (2). The energy density was first computed for each directional component then summed to obtain the overall dispersion relations. In Fig. 5, $k$ is normalized with the Brillouin-zone boundary length $\pi/a$. In the current case with an armchair SWNT, a unit cell is an armchair-shaped monolayer and hence $a=\sqrt{3}a_{c-c}$, where $a_{c-c}$ is the interatomic distance. The data are discrete due to the finite length of the nanotube and the broadening of the spectral peaks indicates the finite temperature effect and phonon scattering. The features of the dispersion relations of a pure $^{12}C$-SWNT obtained from equilibrium MD simulations shown in Fig. 5(a) [22-24] agree with the ones of the previous models [2, 3]. They also agree well with more recent mechanical model with 1st-2nd neighbor directed bonds and radial bond-bending interactions reported by Mahan and Jeon [25].

In Fig. 5(b-d), the dispersion spectra of 50 nm long SWNT isotope-superlattices subjected to quasi-stationary temperature gradients are shown for $\Delta z=0.5$, 1.5 and 5.0 nm, which correspond to 2, 6 and 20 monolayers. Each case represents the regime with dominant zone-folding effect ($\Delta z<\Delta z_{cr}$), crossover ($\Delta z\sim\Delta z_{cr}$) and dominant thermal boundary resistance ($\Delta z>\Delta z_{cr}$). The figures are drawn to provide close-ups of the low frequency and wavevector regime capturing the key phonon branches for heat conduction; LA, TW and F together with four low frequency optical phonon branches ($LO_j$ and $TO_j$). As denoted with dashed lines, the periodicity imposed by the superlattices gives rise to the mini-Brillouin zone with length $\pi/\Delta z$ in $k$-space normalized by the width of the original Brillouin-zone boundary of a pure $^{12}C$-SWNT. For cases with $\Delta z=2a=0.5$ nm [Fig. 5(b)] and $\Delta z=6a=1.5$ nm [Fig. 5(c)], the zone-folding effects are visualized as the local symmetries of the branches with respect to the mini-Brillouin zone-boundaries. The appearance of the mini-bandgaps at the zone boundaries [11] and the corresponding reduction of the group velocity due to the zone-folding effect can be observed. Note that there are distinct discrete jumps of the phonon branches at the intersections with the mini-Brillouin zone boundaries, and the group velocity is locally reduced in the regime near the zone-boundaries. Fig. 5(d) depicts that, in case of $\Delta z=20a=5$ nm, each mini-bandgap becomes smaller, if not diminished, compared with the previous cases, which confirms the switch of the dominant effect from the zone-folding to the thermal boundary resistance. These observations of dispersion relations agree with Simkin and Mahan [10], where the recovery of thermal conductivity in $\Delta z > \Delta z_{cr}$ is characterized by the gradual closing of the mini-bandgaps with increasing $\Delta z$.

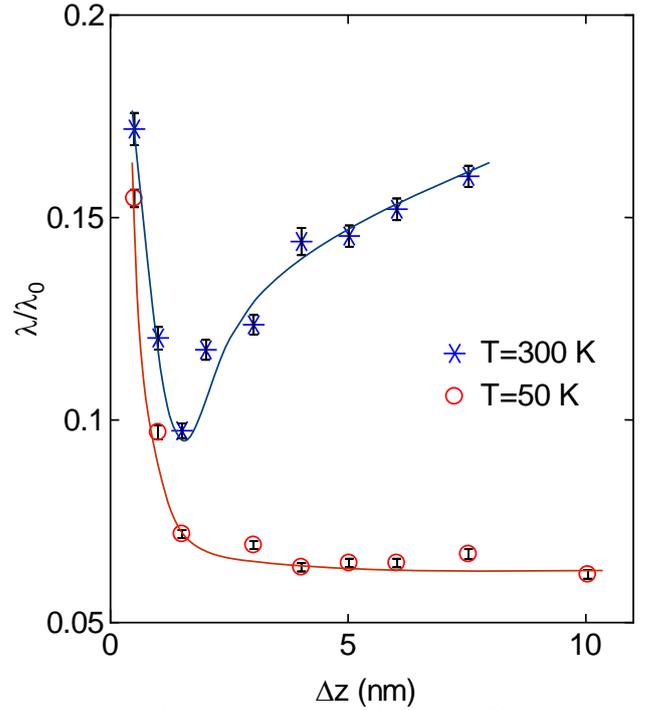

Fig. 4. (Color online) Dimension-less effective thermal conductivity for $^{12}C/^{24}C$ and various period thicknesses $\Delta z$. Thermal conductivity for the pure $^{12}C$-SWNT $\lambda_0$ is 310 W/mK and 1167 W/mK at average temperature of 300 K and 50 K, respectively. The lines are drawn only for eye. The error bars are calculated based on the fitting residuals of the temperature profiles.

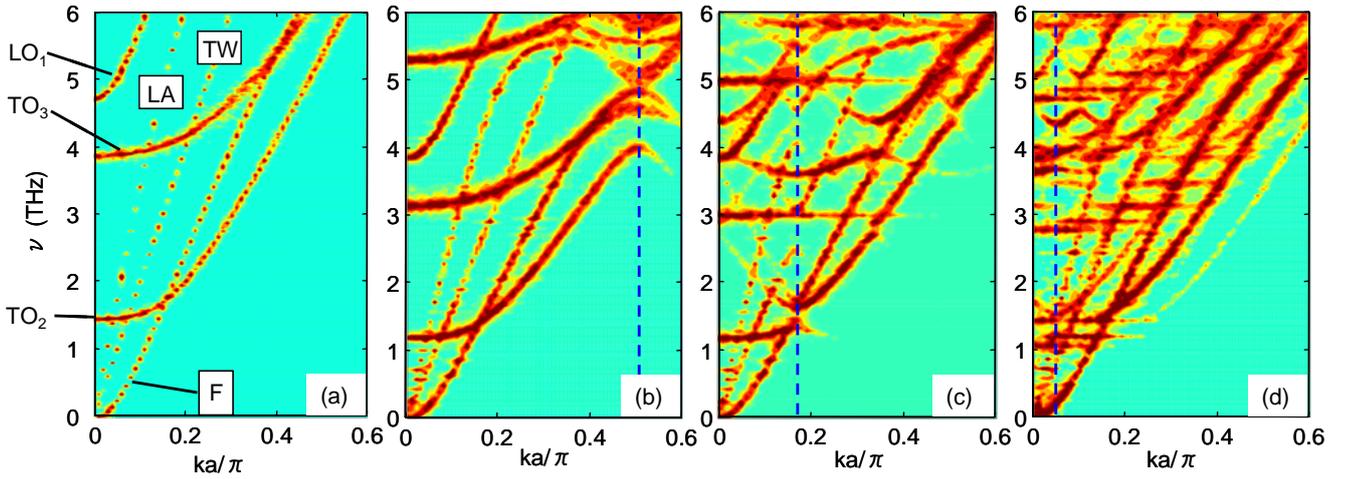

Fig. 5. (Color online) Energy density spectra indicating the dispersion relations of the superlattice SWNTs for the $^{12}C/^{24}C$ system at 300 K. For comparison, the figure (a) shows the dispersion relations of a pure 25 nm long $^{12}C$-SWNT at equilibrium (T=300 K). Figures (b-d) correspond to the non-equilibrium cases with, from left to right, $\Delta z$=0.5, 1.5 and 5nm. Width of the mini-Brillouin zones are denoted by the dashed lines. LA, TW and F indicate the longitudinal acoustic, twisting acoustic and flexure modes. LO and TO indicate longitudinal and transverse optical modes. The subscript denotes the circumferential wave number.

## IV. Conclusions

Heat conduction of SWNT isotope-superlattices was investigated by means of non-equilibrium molecular dynamics simulations. Superlattice structures were constructed by alternately connecting (5, 5)-SWNTs with different masses. At room temperature, implementation of $^{12}C/^{13}C$ superlattice structures can attenuate the thermal conductivity to approximately 56% of the value of the pure $^{12}C$-SWNT. By performing the simulations over a range of period thicknesses, the thermal conductivity was found to take a minimum value at a certain critical period thickness, which reflects the crossover of the zone-folding and the thermal boundary resistance effects. As a consequence, the reduction of the thermal conductivity of superlattice structures beats that of the one-dimensional alloy structure, though the reduction of thermal conductivity is less than the value obtained by two-dimensional random isotope mixing. The results demonstrate that the minimum thermal conductivity appears even in the system dominated by ballistic phonon transport, depending on the distribution of diffusive-ballistic phonons whose mean free path $l_i$ is in the range of $0<l_i<L$. Finally, by calculating the two-dimensional energy density spectra, the zone-folding effects were observed together with the mini-bandgaps, where the opening of mini-bandgaps increased with $\Delta z$ when $\Delta z < \Delta z_{cr}$.

**Acknowledgments**

This work is supported in part by the Japan Society for the Promotion of Science for Young Scientists #1610109 and Grants-in-Aid for Scientific Research #17656072.

**References**

1.  S. Iijima and T. Ichihashi, Nature 363, 60 (1993).
2.  M. S. Dresselhaus, G. Dresselhaus and P. C. Eklund, Science of Fullerenes and Carbon Nanotubes, Academic Press, New York (1996).
3.  R. Saito, G. Dresselhaus and M. S. Dresselhaus, Physical Properties of Carbon Nanotubes, Imperial College Press, London (1998).
4.  S. Maruyama, Y. Taniguchi, Y. Igarashi and J. Shiomi, "Anisotropic Heat Transfer of Single-Walled Carbon Nanotubes" (submitted to J. Therm. Sci. Tech.)
5.  C. Colvard, T. A. Gant, M. V. Klein, R. Merlin, R. Fischer, H. Morkoc and A.C. Gossard, Phys. Rev. B 31, 2080 (1985).
6.  G. Chen, Phys. Rev. B 57, 14958 (1998).
7.  R. Venkatasubramanian, Phys. Rev. B 61, 3091 (2000).
8.  R. Venkatasubramanian, E. Siivola, T. Colpitts and B. O'Quinn, Nature 413, 587 (2001).
9.  B. C. Daly, H. J. Maris, K. Imamura and S. Tamura, Phys. Rev.B 66, 024301 (2002)
10. M. V. Simkin and G. D. Mahan, Phys. Rev. Lett. 84, 927 (2000).

11. S. Y. Ren and J. D. Dow, Phys. Rev. B 25, 3750 (1982).
12. W. S. Capinski, H. J. Maris, T. Ruf, M. Cardona, K. Ploog and D. S. Katzer, Phys. Rev. B 59, 8105 (1999).
13. Y-M. Lin and M. S. Dresselhaus, Phys. Rev. B 68, 075304 (2003).
14. Y. Chen, D. Li, J. Yang, Y. Wu, J. R. Lukes and A. Majumdar, Physica B 349, 270 (2004)
15. C. Dames and G. Chen, J. Appl. Phys. 95, 682 (2004).
16. J. Shiomi and S. Maruyama, "Molecular dynamics of phonon transport in finite-length single-walled carbon nanotubes" (to be submitted).
17. N. Mingo and D. A. Broido, Nano Lett. 5, 1221 (2005).
18. Y. Miyauchi and S. Maruyama, Phys. Rev. B 74, 035415 (2006).
19. D. W. Brenner, Phys. Rev. B 42, 9458 (1990).
20. Y. Yamaguchi and S. Maruyama, Chem. Phys. Lett. 286, 336 (1998).
21. J. Tersoff, Phys. Rev. Lett. 56, 632 (1986).
22. S. Maruyama, Physica B 323, 272 (2002).
23. S. Maruyama, Micro. Therm. Eng. **7**, 41 (2003).
24. J. Shiomi and S. Maruyama, Phys. Rev. B 73, 205420 (2006).
25. G. D. Mahan and G. S. Jeon, Phys. Rev. B 70, 075405 (2004).